\documentclass[12pt]{article}

\usepackage{lscape}
\usepackage{multirow}
\usepackage{bm}
\usepackage{amssymb,amsmath}
\usepackage{amstext}
\usepackage{amscd}
\usepackage{graphics}
\usepackage{epsfig}
\usepackage{shadow}
\usepackage[all]{xy}
\usepackage{hyperref}

\def\wt{\widetilde}
\newcommand{\Rho}{{\mbox{\sf P}}}

\def\Real{\mathbb R}

\newcommand{\eqn}[1]{(\ref{#1})}

\def\d{\partial}

\def\be{\begin{equation}}
\def\ee{\end{equation}}
\def\beq{\begin{equation}}
\def\eeq{\end{equation}}
\def\bea{\begin{eqnarray}}
\def\eea{\end{eqnarray}}

\def\nn{\nonumber}

\def\sideremark#1{\ifvmode\leavevmode\fi\vadjust{\vbox to0pt{\vss
 \hbox to 0pt{\hskip\hsize\hskip1em
 \vbox{\hsize3cm\tiny\raggedright\pretolerance10000
  \noindent #1\hfill}\hss}\vbox to8pt{\vfil}\vss}}}

\begin{document}
\thispagestyle{empty}

\vspace{.8cm}
\setcounter{footnote}{0}
\begin{center}
\vspace{-50mm}
{\Large
 {\bf Gravity, Two Times, Tractors, \\ Weyl Invariance and Six Dimensional Quantum Mechanics}\\[4mm]

 {\sc \small
    R.Bonezzi$^{\mathfrak B}$,  E.~Latini$^{\mathfrak L}$  and A.~Waldron$^{\mathfrak W}$}\\[3mm]

{\em\small
                      ~${}^\mathfrak{B}$ 
Dipartimento di Fisica,\\ Universit\`a di Bologna, via Irnerio 46, I-40126 Bologna, Italy   \\
and INFN sezione di Bologna, via Irnerio 46, I-40126 Bologna, Italy\\
{\tt bonezzi@bo.infn.it}\\[1mm]
          ~${}^\mathfrak{L}$  
          INFN, Laboratori Nazionali di Frascati, CP 13, I-00044 Frascati,~Italy\\
 {\tt latini@lnf.infn.it}\\[1mm]
 
        ${}^{\mathfrak W,\mathfrak L}\!$
            Department of Mathematics\\ 
            University of California,
            Davis CA 95616, USA\\
            {\tt emanuele,wally@math.ucdavis.edu}
   
            }
 }

\bigskip

{\sc Abstract}\\[1mm]

\end{center}

\noindent
Fefferman and Graham showed some time ago that
four dimensional conformal geometries could be
analyzed in terms of six dimensional, ambient,  Riemannian
geometries admitting a closed homothety. 
Recently it was shown how conformal geometry provides
a description of physics manifestly invariant under local choices of
unit systems. Strikingly, Einstein's equations are then equivalent
to the existence of a parallel scale tractor (a six component vector 
subject to a certain first order covariant constancy condition at every point in four dimensional spacetime).
These results suggest a six dimensional description of four dimensional physics, a viewpoint promulgated by 
the two times physics program of Bars. The Fefferman--Graham construction relies on a triplet of operators
corresponding, respectively to a curved six dimensional light cone, the dilation generator and the Laplacian. These form an~$\frak{sp}(2)$ algebra which Bars employs as a first class algebra of constraints in a six-dimensional gauge theory.
In this article four dimensional gravity is recast in terms of six dimensional quantum mechanics
by melding the two times and tractor approaches. This ``parent'' formulation of gravity is built from an infinite set of six dimensional
fields. Successively integrating out these fields yields various novel descriptions of gravity including a new four dimensional one
built from a scalar doublet, a tractor vector multiplet and a conformal class of metrics.

\newpage

\tableofcontents


\section{Introduction}
Theories with extra dimensions have been heavily scrutinized since the time of Kaluza and Klein~\cite{KK}.
The terminus of this train of thought is String Theory which attempts to encode the couplings of four dimensional 
theories in the geometry of hidden higher dimensions. A simpler and more generic rationale for further dimensions,
however, might follow a line of reasoning similar to Einstein's original identification of time as an additional coordinate,
along with a gauge principle---general coordinate invariance---guiding the construction of physical theories in terms of
Riemannian geometry.

In this article, we focus on two fairly recent suggestions that physics is inherently six dimensional.
Firstly, motivated by duality and holographic arguments, Bars observed that many seemingly different four dimensional
particle models could be regarded as gauge fixed versions of a single underlying six dimensional model. 
In fact the idea of using six dimensions to describe four dimensional physics dates back to Dirac~\cite{Dirac:1936fq}.
What is notable about Bars' ``two times physics'' \cite{Bars:1997bz} (see~\cite{Bars:2010xi} for an overview) is that it aims ultimately to describe {\it any} physical system, whereas
Dirac's work pertained only to models with conformal symmetry\footnote{In fact there is a extensive literature on the handling of four dimensional conformal theories using six dimensional methods. Pertinent contributions include Boulanger's conformal tensor calculus~\cite{Boulanger:2004eh}, the conformal space method of~\cite{Preitschopf:1998ei}, the conformal higher spin studies~\cite{Shaynkman:2004vu}, the BRST conformal parent action method of~\cite{Bekaert:2009fg}, and the application to scattering amplitudes in~\cite{Weinberg}.}.

The second approach relies on replacing Riemannian geometry with conformal geometry so that physics is described by conformal classes 
of metrics and all equations are manifestly locally Weyl invariant. This is achieved by utilizing the simple physical principle that
no physical quantity can depend on local choices of unit system which implies there must exist a way to write any physical system
in a Weyl invariant way~\cite{mass,Shaukat:2010vb}. Weyl invariance is intimately related to conformal symmetry, and for reasons very similar to those
first observed by Dirac, manifest Weyl invariance can be achieved by grouping existing four dimensional physical quantities in
six dimensional multiplets known as ``tractors''. This approach relies heavily on tractor calculus~\cite{Thomas,tractommathreview,GP}, a mathematical machinery designed
for efficiently handling conformal geometries. Not only does the tractor approach identify a simple gauge principle---local 
unit invariance---for constructing models, it also identifies the additional timelike coordinate in two times physics as the choice of scale.

In this article we map out the relationship between the two times and tractor approaches, since they are in fact highly complementary,
and in doing so  present seven different formulations of four dimensional Einstein gravity\footnote{Our results are valid for any spacetime dimensionality, and all formul\ae~will be presented as functions of $d$, the spacetime dimension. We will, however often use the shorthand ``four'' to stand for $d$-dimensional and ``six'' to stand for $(d+2)$-dimensional.}, several of which are novel; they are summarized by the action principles~(\ref{conformally impr scalar}, \ref{I^2 action for gravity}, \ref{BRST}, \ref{action}, \ref{action2},  \ref{action3}, \ref{quantum}).
Of these, the action~\eqn{BRST} can be viewed as a parent action depending on infinitely many fields living in a six dimensional spacetime
while all other theories are gauge fixed versions of this parent action. This starting point was first proposed by Bars as part of his
two times description of physics although not precisely as a four dimensional theory of gravity~\cite{Bars:2006ir}. 
This action comes from a BRST
quantization of the worldline conformal group gauge symmetries of a two times particle model\footnote{Massless four dimensional spinning particles were obtained earlier from six dimensions by Siegel in~\cite{Siegel:1988ru} and further studied in~\cite{Kuzenko:1995mg}. }. The operators generating local worldline
conformal transformations form the gravity multiplet of the model. Bars' action couples this gravity multiplet to a scalar multiplet which
can be viewed as a dilaton. This fits extremely well with the tractor description of gravity in terms of a conformal class of metrics
coupled to a scale field---the gauge field for local changes of unit systems.

There is an alternative proposal for a two times description of four dimensional gravity due to Bars~\cite{Bars:2008sz}. It has the advantage that at least 
part of the equations for the generators of worldline conformal transformations follow from an action principle. On the other hand,
unlike the action~\eqn{BRST}, it does not make the worldline conformal group $\frak{sp}(2)$ symmetry---a central component of
the two times set-up---manifest. It turns out that the two approaches are in fact equivalent, a fact that follows rapidly using tractor technology.

The tractor approach takes standard four dimensional physical quantities and groups them in Weyl-multiplets labeled by $SO(d,2)$ representations\footnote{For example, for a relativistic particle, from the four-velocity $v_\mu$, the component of the four-acceleration $a^\mu$ and  the vanishing function, one can build a tractor ``six-velocity'' $V_M=(\frac{v.a}{v.v},e^\mu{}_m v_\mu,0)$ transforming as a multiplet under Weyl transformations according to~\eqn{tractor gauge transformation}.} known as tractors. These tractors are functions 
of four dimensional spacetime. In particular, from the scale field~$\sigma$ (the spacetime dependent Planck's constant), one builds a
tractor vector $I^M$ known as the scale tractor. Like any tractor, under Weyl transformations it undergoes a tractor gauge transformation
which in turn defines a covariant derivative known as the tractor connection\footnote{In fact, the tractor connection also appears in the Yang--Mills-like construction of conformal supergravity~\cite{Kaku:1977pa}.}~\cite{tractommathreview}. The beauty of this approach is that the Einstein condition
amounts to the scale tractor being parallel with respect to this connection. The length of the scale tractor is therefore parallel for
physical geometries and in fact measures the cosmological constant. Upon coupling to matter, it also provides a massive coupling constant.
Remarkably, even though the small size of the cosmological constant might seem to make the length of the scale tractor inappropriate
for setting particle physics mass scales, including backreaction immediately solves this ``cosmological constant hierachy problem''~\cite{Bonezzi:2010ky}.
In fact, parallel scale tractors form the first part of a link between
the tractor and two times descriptions of gravity.

The link between two times physics and tractors is completed by the ambient formulation of tractor calculus developed by~\cite{CG,GP,Gover:2009vc}.
The main idea underlying ambient tractors relies on the Fefferman--Graham description of four dimensional conformal geometries in terms
of six dimensional Ricci flat geometries admitting a closed homothety~\cite{FG}. The latter condition implies that the six dimensional ambient 
geometry enjoys a curved null cone with a dilation-like vector field. This allows four dimensional conformal geometries to be realized
as rays in this ambient lightcone. Bars' $\frak{sp}(2)$ triplet of worldline conformal group Noether charges can be viewed, respectively,  as the 
defining function for the ambient null cone, dilation generator and the harmonic condition obeyed by the Weyl tensor for a Ricci flat
geometry. Essentially taking the old Fefferman--Graham ambient metric construction, alongside with the idea of describing
unit invariant four dimensional physics with conformal geometry leads one directly to Bars' two times physics program.
Needless to say, this confluence of mathematical and physical technologies is likely to lead to major advances in both fields.

Our paper is organized as follows:
In section~\ref{Scale Tractor} we review how Einstein gravity can be recovered in the tractor framework as a parallel condition on the scale tractor, and we fix conventions and notations. In particular we define the tractor connection and we introduce the main tractor operators.
In section~\ref{ambient}, we  set out the ambient description of  tractors and introduce the triplet of $\frak{sp}(2)$  operators
underlying the two times approach. We discuss the latter in detail in section~\ref{2TP}, where we introduce the most general deformation of the flat $\frak{sp}(2)$ algebra which contains an infinite tower of background fields. 
In section~\ref{main} we give the main new results based on a detailed analysis of Bars' BRST parent field theory action. 
By careful gauge choices and identification of the dilaton field we produce the slew of new descriptions of four dimensional gravity mentioned
above as well as establishing the link between tractor and two times approaches. In appendix~\ref{alt} we give a succinct tractor analysis   
of Bars' alternate proposal for a two times gravity theory. In our conclusions (section~\ref{conc}) we discuss the six dimensional
quantum mechanical origin of four dimensional gravity, a candidate master theory generating the~$\frak{sp}(2)$ and dilaton dynamics,
a frame-like formulation of two times physics and the relation between the towers of auxiliary fields  of the two times approach and an unfolding of the full (non-linear) four dimensional Einstein's equations.

\section{Gravity and Parallel Scale Tractors}\label{Scale Tractor}


It is well known that the Einstein--Hilbert gravitational action
 can be
viewed as the gauge fixed version of a conformally improved scalar
field theory~\cite{Zumino,Deser0} 
\begin{equation}\label{conformally impr scalar}
S[\varphi,g]=-\scalebox{1.1}{$\frac{4(d-1)}{(d-2)}$}\int
d^dx\sqrt{-g}\Big[\frac12\,(\nabla\varphi)^2+\frac18\ \frac{d-2}{d-1} \ R \ \varphi^2
\Big]\, ,
\end{equation}
which is invariant under local Weyl rescalings 
$\Omega(x)$, transforming $\varphi\mapsto\Omega^{\frac{2-d}{2}}\varphi$ and
\be\label{Weyl}
g_{\mu\nu}\mapsto\Omega^2g_{\mu\nu}\, .
\ee
On the one hand this seems a rather trivial observation because
choosing the gauge in which
$\varphi$ is constant and equal to $\kappa^{-1}$, one recovers the usual
gravity action $S(g,\kappa^{-1})=-\frac{1}{2\kappa^2}\int d^dx\sqrt{-g}R$.
To see that this is in fact a statement of fundamental importance, first note that
the Weyl transformation~\eqn{Weyl} defines the equivalence class relation $g_{\mu\nu}\sim \Omega^2 g_{\mu\nu}$
of a conformal class of metrics~$[g_{\mu\nu}]$, so that physics can be cast in terms 
of conformal, rather than Riemannian geometry. Secondly, note that the Weyl transformation~\eqn{Weyl}
amounts to making local redefinitions of unit systems, which along with general coordinate invariance, is a 
symmetry that any formulation of physics must enjoy.

So far there is no hint of any six dimensional quantities. To see these, we attempt to write the Weyl invariant
formulation~\eqn{conformally impr scalar}  of Einstein--Hilbert gravity as the square of a single vector~$I^M$
\begin{equation}\label{I^2 action for gravity}
S[g,\sigma]=\scalebox{1.1}{$\frac{d(d-1)}{2}$}\int d^dx\frac{\sqrt{-g}}{\sigma^d}\, I^M I_M\, .
\end{equation}
The six component vector
\begin{equation}\label{scaletractor}
I^M
= \begin{pmatrix}\sigma\\[3mm]
\nabla^m\sigma\\[1mm]
-\frac1d\Big[\Delta+\Rho\Big]\sigma\end{pmatrix}\;,
\end{equation}
is called the ``scale tractor'' and is distinguished by its transformation properties under Weyl transformations. Here the scalar $\sigma=\varphi^{\frac2{2-d}}$ is simply a relabeling of the dilaton~$\varphi$ so that it has unit Weyl weight
$$
\sigma\mapsto \Omega\, \sigma\, .
$$
The field $\sigma$ is often called the ``scale'' since it measures the relative choice of unit system from point to point in spacetime.
Also, it is often convenient to work with the Schouten tensor $\Rho_{\mu\nu}$ which is the trace adjusted Ricci-type
tensor, defined by
\begin{equation}\nn
\Rho_{\mu\nu}=\frac{1}{d-2}\Big(R_{\mu\nu}-\frac{1}{2(d-1)}g_{\mu\nu}R\Big)\;,
\end{equation}
and its trace is denoted $\Rho=\Rho_\mu^\mu$.

The main features of the action~\eqn{I^2 action for gravity} are
\begin{itemize}
\item It depends on conformal classes of metrics, embedded in
the double equivalence class $[g_{\mu\nu},\sigma]\sim[\Omega^2g_{\mu\nu},\Omega\sigma]$. 
This allows for manifest Weyl invariance while still specifying a canonical metric
$g_{\mu\nu}^0$ in the conformal class satisfying
$[g_{\mu\nu},\sigma]\sim[g_{\mu\nu}^0,\kappa^{\frac{2}{d-2}}]$. 
\item The measure $\sqrt{-g}\, \sigma^{-d}$ is separately Weyl invariant, as is also
the square of the scale tractor $I^2$. This holds because the scale tractor $I^M$
transforms under particular local $SO(d,2)$ transformations known as tractor
gauge transformations.
\item Einstein's equations amount to the scale tractor being parallel with respect to the tractor
connection, exactly the covariant derivative implied by tractor gauge transformations.
\item The ``length'' of the scale tractor  measures the cosmological constant. Hence Ricci flatness
 implies  a lightlike  scale tractor.
\end{itemize}
Let us explain these points and the key ingredients of tractor calculus in more detail.

  From the  four dimensional viewpoint, a six-component multiplet $(V^+$, $V^m$, $V^-)$
with $m=0,..,d-1$, forms a weight $w$ tractor vector $V^M$, $M=+,m,-$,
if under Weyl transformations it obeys the tractor gauge transformation :
\be\label{tractor gauge transformation}
V^M \mapsto \Omega^w U^M{}_N V^N\, , \qquad
U^M{}_N=
\begin{pmatrix}
\Omega & 0 & 0 \\[2mm]
\Upsilon^m & \delta^m_n&0 \\[2mm]
-\frac{\Upsilon^2}{2\Omega} & -\frac{\Upsilon_n}\Omega &\frac1\Omega
\end{pmatrix}\, ,
\ee
where $\Upsilon_\mu = e_\mu{}^m \Upsilon_m =\Omega^{-1}\partial_\mu
\Omega$. In section~\ref{ambient} we will see that tractors naturally live as six-vectors in
a six dimensional, signature $(4,2)$
 spacetime endowed with a curved light-cone structure. The
reduction to four dimensions  induces a tractor-covariant
connection:
\begin{equation}\label{tractorconnection}
{\cal D}_\mu = \begin{pmatrix}\partial_\mu&-e_{\mu n} & 0 \\[1mm]
    \Rho_\mu{}^m
&\nabla_\mu{}^m{}_n&e_\mu{}^m\\[1mm]0&-\Rho_{\mu
n}&\partial_\mu\end{pmatrix}\, ,
\end{equation}
such that $${\cal D}_\mu V^M\mapsto \Omega^wU^M{}_N\big[{\cal D}_\mu +w\Upsilon_\mu\big]V^N\, .$$
By means of the tractor connection one can construct a weight $-1$
tractor-vector operator, the so called ``Thomas $D$-operator'', which acting on weight~$w$
tractors reads:
\begin{equation}\label{thomasd}
D^M=
\begin{pmatrix}
w(d+2w-2)\\[2mm]
(d+2w-2) {\cal D}^m\\[2mm]
-({\cal D}_\mu{\cal D}^\mu+
w\Rho)
\end{pmatrix}\;.
\end{equation}
Acting with the Thomas $D$-operator on the scale~$\sigma$, we obtain a weight~$0$ tractor-vector, 
the scale tractor 
$$I^M=\frac1d D^M\sigma\, ,$$ which has components
exactly given by~\eqn{scaletractor}.

The scale tractor's main importance is twofold: first, in tractor theories
it controls the coupling of matter to scale in a Weyl-covariant way~\cite{mass},
parametrizing the breaking of local scale invariance in the
$\sigma=constant$ physical gauge. On the other hand, $I^M$ is
closely related to gravity itself: remarkably, the gravity-dilaton
action (\ref{conformally impr scalar}), can be
written entirely in terms of the scale tractor as in~\eqn{I^2 action for gravity}
where tractor indices are raised and lowered with the $SO(d,2)$
invariant metric
$$
\eta_{MN}=\begin{pmatrix}0&0&1\\[1mm]0&\eta_{mn}&0\\[1mm]1&0&0\end{pmatrix}\;.
$$

To see that a tractor-parallel scale tractor, \emph{i.e.} $
{\cal D}_\mu I^M=0\,,$
amounts to Einstein's equations
we explicitly compute the tractor
derivative of $I^M$ that, once evaluated at the choice of constant scale
$\sigma=\sigma_0$, reads
\begin{equation}\label{parallel scale tractor}
\mathcal{D}_\mu I^M\rvert_{\sigma=\sigma_0}=\sigma_0
\begin{pmatrix}0\\[3mm] \Rho_\mu{}^m-\frac1d
e_\mu{}^m\Rho\\[3mm]-\frac1d\d_\mu\Rho\end{pmatrix}\;.
\end{equation}
Setting this to zero  says $R_{\mu\nu}=\frac1d
g_{\mu\nu}R$ and $R=constant$, so that $g_{\mu\nu}$ is precisely an Einstein
manifold. This happens at the choice of scale $\sigma=\sigma_0$, so we
can say that the scale tractor is parallel when the metric is
conformally Einstein:
$$
\mathcal{D}_\mu I^M=0\quad\Leftrightarrow\quad
g_{\mu\nu}=\Omega^2g_{\mu\nu}^0\quad\text{with}\quad
R_{\mu\nu}(g^0)\propto \,g_{\mu\nu}^0\; .
$$
Moreover, if
the scale tractor is parallel then its length squared  $I^2\equiv I^MI_M$ is constant, and
 proportional to the cosmological constant.
 
 Geometrically  the scale tractor can be viewed as coming from a vector perpendicular to a hypersurface in six dimensions.
 The intersection of that hypersurface with a (curved) lightcone defines a conformal class of metrics on the four dimensional
 intersection. This picture relies on a six dimensional ambient description of tractors which we describe in the next section. 
 Given the significance of the scale tractor $I^M$, it would be extremely interesting to formulate four dimensional gravity 
 in terms of an independent six component vector field. That result is obtained by combining ambient tractors with Bars' two times physics
 proposal and is given in section~\ref{main}.





\section{Ambient Tractors }
\label{ambient}

The importance of six-dimensional spacetimes for describing conformally invariant four-dimensional theories has been clear
since the work of Dirac~\cite{Dirac:1936fq}. (Perhaps the simplest motivation for this is that the Minkowski space conformal group $SO(4,2)$ 
acts naturally on the flat Lorentzian space $\Real^{4,2}$.) Weyl invariance ensures rigid conformal symmetry whenever the metric enjoys conformal
isometries; this suggests that four-dimensional conformal geometries can be studied in terms of six-dimensional Riemannian geometries.
This was shown to be the case by Fefferman and Graham~\cite{FG} who formulated the problem of constructing conformal invariants
in terms of a six-dimensional ambient metric. This idea was extended to the tractor calculus description of conformal geometry in the series of articles~\cite{CG,GP} (see also~\cite{Gover:2009vc}).

Based on duality and holographic arguments, the two times approach of Bars advocates that four dimensional physics (irrespective of whether it enjoys rigid conformal symmetry or not) can be described using a six dimensional spacetime. 
The tractor approach of Gover {\it et al} uses the simple principle of invariance under local choices of unit system to argue
that four dimensional physics should be formulated in terms of conformal geometry. Since the latter, in turn, enjoys an ambient
six dimensional formulation, local unit invariance and tractors also support a formulation of four dimensional physics using a six dimensional 
spacetime. In this section we give the main ingredients of the six dimensional ambient description of tractor calculus.

A four dimensional conformal manifold equipped with an equivalence class of metrics $[g_{\mu\nu}]$,
with equivalence defined by local Weyl transformations
$$
g_{\mu\nu}\mapsto \Omega^2 g_{\mu\nu}\, ,
$$
can be viewed as the space of rays in a five dimensional null hypersurface embedded in a six dimensional
Riemannian ambient space with metric $G_{MN}$. 
Specializing to the conformally flat case, consider the ambient space  $\Real^{4,2}$ with the standard flat Lorentzian metric~$
dX^M \eta_{MN} dX^N
$,
which enjoys a closed (and therefore hypersurface orthogonal) homothety given by the dilation/Euler operator
$
X^M\frac{\partial}{\partial X^M}.
$
The zero locus of the homothetic potential
$
X^M X_M\equiv  X^2$
defines a five dimensional  null cone 
so the space of null rays~$\xi^M$ subject to the equivalence relation
$\xi^M\sim \Omega\,  \xi^M$ (where $\Omega\in \Real^+$)
is four dimensional and determines a (conformally flat) four dimensional conformal structure.
The conformal class of metrics follows  by letting 
$\xi^M(x)$ be a  section of the null cone. 
The ambient metric then pulls back to a four dimensional metric $ds^2=d \xi^M d\xi_M$. 
Choosing a different section $\xi^M(x)$ results in a conformally related metric.
For example, in the conformally flat setting, de Sitter, Minkowski and anti de Sitter space
 all inhabit the same conformal class. In this case the tractor connection of~\eqn{tractorconnection} is the pullback of the 
 Cartan--Maurer form of $SO(4,2)$ to the conformally flat four dimensional space time described as a coset $SO(4,2)/P$
 where $P$ stabilizes a lightlike ray.
 
 The above flat model of conformal geometry, as the space of lightlike rays in a six dimension ambient space, 
 extends to curved spaces and general conformal structures as follows:
 A four dimensional conformal structure determines a Fefferman--Graham ambient metric which admits a hypersurface orthogonal
homothety. In the flat case this homothety is generated by the Euler vector field whose components coincide with the standard Cartesian coordinates. In the curved ambient construction,
the corresponding homothetic vector field will still be denoted by $X^M$ (which are {\it not}
generally coordinates for which we reserve the notation $Y^M$).
The key identity is then the equation
\be
G_{MN}=\nabla_M X_N\, ,
\label{key}\ee
where $G_{MN}$ is the ambient metric and $\nabla$ is its Levi-Civita covariant derivative.
This condition already suffices to uniquely determine a four dimensional conformal structure.
The symmetric part of \eqref{key} implies the homothetic conformal Killing equation
while its antisymmetric part says that the one-form dual to $X^M$ is closed. Indeed this one form is
exact
$$
X_M=\frac12 \nabla_M X^2\,   .
$$
Clearly, the ambient metric is the double gradient of the homothetic
potential $G_{MN}=\frac12 \nabla_M \partial_N X^2$.
The zero locus of the potential $X^2$ defines a curved cone,  a quotient of which recovers  
the four dimensional conformal manifold.
Observe that the above identities for the ambient metric imply
\be
X^M R_{MNRS}=0=(X^T \nabla_T + 2) R_{MN}{}^R{}_S\, .
\nn\label{Xriem}
\ee
To ensure uniqueness of the ambient metric for a given four dimensional structure, Fefferman and Graham
require that the ambient
metric is formally  Ricci flat in any odd dimension (to all orders), and Ricci flat to finite order
 in the defining function~$X^2$ in even dimensions greater than or equaling four. For our purposes, uniqueness of the underlying
 four dimensional conformal structure is all we need, so we will typically work with six dimensional ambient metrics subject to~\eqn{key}
 but need not impose six dimensional Ricci flatness.

The Rosetta Stone between six dimensional ambient space operators and the
Thomas $D$-tractor operator~\eqn{thomasd} on a four dimensional conformal manifold was first given in~\cite{GP}and simply reads
\bea
D_M\ &\equiv&\nabla_M (d+2X^N\nabla_N-2)-X_M\, \Delta\, .\label{DDD}
\eea
The canonical tractor of~\cite{tractommathreview} corresponds to the vector field $X^M$ while tractor weights are eigenvalues of the operator $X^M\nabla_M$.
(In~\cite{Gover:2009vc}, it was realized that these operators are related to a momentum space representation of the ambient space conformal group.)
Tractor tensors $T^{M_1\ldots M_s}(x)$ (sections of weighted tractor tensor bundles over four dimensional spacetime) can then be viewed as equivalence classes of six dimensional ambient space tensors
\be
T^{M_1\ldots M_s}(Y) \sim T^{M_1\ldots M_s}(Y) + X^2 \ U^{M_1\ldots M_s}(Y) \, ,\label{cone}
\ee
subject to a weight constraint
\be
X^M \nabla_M T^{M_1\ldots M_s} = w\,  T^{M_1\ldots M_s}\, .
\label{w}\ee
The equivalence relation can also be handled by working with weight $w-2$ ambient space tensors of the form $$\delta(X^2)\ T^{M_1\ldots M_s}\, ,$$ subject to the constraint $X^2=0$.
It is not difficult to check that the ambient operator~\eqn{DDD} is well defined on equivalence classes defined by the cone condition~\eqn{cone}.

The equivalence relation~\eqn{cone} and weight constraint~\eqn{w} do not define a unique extension of a four dimensional tractor 
to the six dimensional ambient space. For that, one needs to ``fix a gauge'' for the equivalence relation. A convenient choice is to require 
that six dimensional quantities are harmonic. The first example of this is the Ricci flat condition of Fefferman--Graham (because the remaining Weyl part of the ambient Riemann curvature is then harmonic). In fact, it is easily verified that the triplet of operators
\be
\{X^2\, , X^M\nabla_M + \frac{d+2}{2}\, , \Delta\}\, ,  \label{sp2}
\ee
obey an $\frak{sp}(2)$ Lie algebra. This algebraic fact underlies Bars' two times approach described in the next section.

\section{Two Times Physics}
\label{2TP}

A simple starting point for understanding two times physics, is the Howe dual pair~\cite{howe}
\be\label{how}
\frak{sp}\big(2(d+2)\big) \supset \frak{sp}(2) \oplus \frak{so}(d,2)\, .
\ee
This Lie algebra statement---namely that $\frak{sp}(2)$ and $\frak{so}(d,2)$ are maximal cocommutants in $\frak{sp}\big(2(d+2)\big)$---says that imposing as constraints an $\frak{sp}(2)$ subalgebra of the natural $\frak{sp}\big(2(d+2)\big)$ algebra acting on a $d+2$ dimensional phase space, leaves a residual $\frak{so}(d,2)$ global symmetry algebra. This latter algebra generates the conformal isometries of $d$-dimensional Minkowski (or more generally conformally flat) spacetime.  

Consider, for example, Bars' approach to the relativistic particle~\cite{Bars:2001xv,Bars:2001uy}. Instead of requiring worldline reparameterization invariance
and therefore a four dimensional Hamiltonian constraint, Bars requires local worldline conformal invariance under $\frak{so}(2,1)\cong
\frak{sp}(2)$ which imposes a triplet of first class constraints. In four dimensions a three dimensional constraint algebra would be too constraining, but as is clear from the Fefferman--Graham ambient space construction described above, if this constraint algebra acts in six dimensions as in~\eqn{sp2}, the null cone and weight constraints perform the reduction to four dimensions leaving a single Hamiltonian
constraint just as in the standard approach. By making different gauge choices for the local $\frak{sp}(2)$ symmetry, one can obtain
a plethora of four dimensional models---``holographic shadows''---all encompassed by a single six dimensional one~\cite{Bars:2000mz}.

The above discussion pertains to single particle models propagating in fixed backgrounds. Our chief interest is a description of
four dimensional field theories and in particular four dimensional gravity. For that, two main ingredients are required. Firstly we must quantize the underlying particle model so that, in turn, quantum mechanical wave functions can be reinterpreted as quantum fields. Secondly
we need to write equations of motion for the background fields. Both steps can be achieved in a unified way by working with quantum mechanical operators. (An alternative approach employed heavily by Bars~\cite{Bars:2001uy,Bars:2001ma} is to employ phase space quantization technology~\cite{book}, but we find working directly with quantum mechanical operators to be more direct.) 

Our model, described in detail in the next section, will be built from two multiplets, the first ``gravity multiplet''  will describe ambiently a conformal class of metrics
along with an additional vector field intimately related to the scale tractor of section~\ref{Scale Tractor}. 
The second ``dilaton multiplet''  describes the dilaton or scale field (or in other words a spacetime-varying Planck's constant). Equations of motion for the gravity multiplet have already been proposed by Bars~\cite{Bars:2001um}. Classically they amount to a triplet of Hamiltonians $Q_{ij}=Q_{ji}$ ($i,j=1,2$) on a $2(d+2)$ dimensional phase space subject to an $\frak{sp}(2)$ algebra under Poisson brackets
\be\label{symp}
\{Q_{ij},Q_{kl}\}=\varepsilon_{kj}Q_{il}+\varepsilon_{ki}Q_{jl}+\varepsilon_{lj}Q_{ik}+\varepsilon_{li}Q_{jk}\, .
\ee
Here one must solve for the $Q_{ij}$ modulo gauge transformations corresponding to canonical transformations
\be\label{cgauge}
Q_{ij}\mapsto Q_{ij} + \{\epsilon,Q_{ij}\}\, .
\ee
An elegant solution has been found by Bars~\cite{Bars:2001um} by choosing Darboux coordinates $\{P_M,Y^N\}=\delta^N_M$, expanding in powers of the momentum $P_M$ shifted by some vector field $A_M(Y)$, and then partially fixing the gauge invariance~\eqn{symp} so that
\be\label{Qcl}
Q=
\begin{pmatrix}
X^M G_{MN}(Y) X^N & X^M \wt P_M \\[3mm] X^M \wt P_M  & \Sigma(Y) + \wt P_M G^{MN}(Y) 
\wt P_N + H(\wt P,Y)
\end{pmatrix}\, ,
\ee
where
\bea
\wt P_M&\equiv&P_M+A_M(Y)\, ,\nn\\[2mm]
H(\wt P,Y)&\equiv&\sum_{k=2}^\infty H^{M_1\ldots M_k}(Y) \wt P_{M_1}\cdots\wt P_{M_k}\, .\nn
\eea
In addition,
this result is intimately connected to  ambient tractors, 
because the algebra~\eqn{symp} requires the metric $G_{MN}$ appearing in~\eqn{Qcl}
to obey the closed homethety condition~\eqn{key}. Moreover the vector field $A_M$ appearing in $\wt P^M$ obeys
\be\label{XF}
X^M F_{MN} \equiv (\pounds_X+1) A_N -\nabla_N(X^MA_M)=0\, ,
\ee
and the scalar $\Sigma$ and totally symmetric tensors $H^{M_1\ldots M_k}$ are subject to weight conditions
\bea
(\pounds_X+2)\, \Sigma\: \: &\equiv&\  (X^M \nabla_M +2)\ \Sigma=0\, ,\nn\\[2mm]
(\pounds_X+2)H^{M_1\ldots M_k}&\equiv& (X^M \nabla_M +2-k) H^{M_1\ldots M_k}=0\, .\label{weight}
\eea
Classically the tensors $H^{M_1\ldots M_2}$ must also be transverse to the homothetic vector field $X^M$.
The above solution still enjoys residual gauge symmetries of the form~\eqn{cgauge}. The beauty of Bars' solution 
is that these residual transformations amount to diffeomorphisms of the tensors $X^M$, $G_{MN}$, $A_M$, $\Sigma$ and
$H^{M_1\ldots M_k}$, abelian Maxwell gauge transformations of $A_M$, as well as a certain class of higher rank symmetries of
the symmetric tensors $H^{M_1\ldots M_k}$ which we will discuss in detail later.

To quantize the Hamiltonians $Q_{ij}$, we look for operators acting on wavefunctions depending on coordinates $Y^M$.
We express these as expansions in the covariant derivatives $\wt \nabla_M=\nabla_M + A_M$. This amounts to a choice of quantum orderings for a basis of all operators acting on wavefunctions. More precisely, momenta $P_M$ act on  wavefunctions
as  derivatives $\partial_M$, but we add subleading ordering terms to higher powers of momenta in order to maintain covariance.
We then require that the quantum commutator of the $Q_{ij}$'s obeys the $\frak{sp}(2)$ algebra
\be\label{QQ}
[Q_{ij},Q_{kl}]=\varepsilon_{kj}Q_{il}+\varepsilon_{ki}Q_{jl}+\varepsilon_{lj}Q_{ik}+\varepsilon_{li}Q_{jk}\, ,
\ee
modulo the quantum symmetry
\be\label{Qs}
Q_{ij}\mapsto Q_{ij} + [\epsilon,Q_{ij}]\, ,
\ee
whose parameter $\epsilon$ is now itself an operator. This system of equations has been proposed by Bars in an equivalent phase space and star product quantization~\cite{Bars:2001um}. Quantization
necessitates a slight modification of Bars' classical solution to 
\be\label{Qqu}
Q=
\begin{pmatrix}
X^2 & X^M \wt \nabla_M +\frac{d+2}{2}\\[3mm] X^M \wt \nabla_M  +\frac{d+2}{2}\ & \ \Sigma +  
\wt \nabla^2 + H(\wt\nabla,Y)
\end{pmatrix}\, ,
\ee
with
$$
H(\wt \nabla,Y)\equiv\sum_{k=2}^\infty H^{M_1\ldots M_k}(Y) \wt \nabla_{M_1}\cdots\wt \nabla_{M_k}\, .
$$
Here the closed homothety, curvature and weight conditions are unaltered from their classical counterparts~(\ref{key},\ref{XF},\ref{weight}), 
but the transverse conditions on the {\it symmetric} tensors $H^{M_1\ldots M_k}$ are modified to read
\be\label{higher}
2X_M H^{MM_2\ldots M_k} + (k+1) H_M{}^{MM_2\ldots M_k}=0\, .
\ee
From this we learn iteratively that the trace of $H^{MN}$ vanishes, the trace of $H^{MNR}$ is the part of $H^{MN}$ parallel to $X^M$ {\it etcetera}. More succinctly, the condition~\eqn{higher} just says
\be\nn
[X^2,H(\wt\nabla,Y)]=0\, .
\ee
But now let us examine which gauge symmetries respect the quantum solution~\eqn{Qqu}.
Firstly, expanding the gauge parameter in powers of $\wt\nabla_M$
$$
\epsilon(\wt \nabla,Y)=-\alpha(Y) + \xi^M(Y) \wt \nabla_M + \varepsilon(\wt \nabla,Y)\, ,
$$
where all terms of quadratic order and higher are stored in $\varepsilon$, it is  easy to verify that the zeroth and first order terms generate abelian gauge transformations
$$A_M\mapsto A_M + \nabla_M \alpha\, ,$$
and diffeomorphisms with parameter $\xi^M$. These are desirable symmetries, so we do not want to gauge fix them at this juncture. We still have the higher order gauge freedoms in~$\varepsilon$, although these are not completely arbitrary:
Requiring $Q_{11}=X^2$ to  be inert, the gauge parameter~$\varepsilon$ obeys the same commutation 
relation with the homothetic potential as $H$
\be\label{XXE}
[X^2,\epsilon]=0\, .
\ee
Furthermore, invariance of $Q_{12}$ implies that $$[X^M \wt\nabla_M,\varepsilon]=0\, .$$
It follows that $\delta Q_{22}\equiv [\varepsilon,\Sigma + \wt\nabla^2+H]$ obeys the same conditions as $H$, namely
$$[X^2,\delta Q_{22}]=0 = [X^M\wt\nabla_M,Q_{22}]+2Q_{22}\, .$$
Now we define a vector
$$U_M\equiv \nabla_M \Sigma\, ,$$
and note that 
\be\label{shift}
[\varepsilon,\Sigma]= \frac 12 \varepsilon^{MN}\pounds_U G_{MN} + 
\varepsilon^{MN} U_M \wt \nabla_N + \sum_{k=3}^\infty\ k\, 
\varepsilon^{M_1\ldots M_k} (U_{M_1}\wt \nabla_{M_2}\cdots \wt \nabla_{M_k})_{\rm W}\, ,
\ee
where $(\bullet)_{\rm W}$ denotes Weyl ordering in the symbols $(U,\wt\nabla)$.

We now make the assumption that the vector $U_M$ is non-vanishing. Certainly, the set of vanishing $U_M$ is measure zero (a situation similar to non-invertible metrics among the space of $4\times 4$ matrices). Bars has
suggested that models with vanishing $U_M$ might describe a novel ``higher spin branch'', but we do not pursue
this line of argument any further here. With $U_M$ non-vanishing the space of rank two and higher symmetric tensors
$U_M \varepsilon^{MM_1\ldots M_k}$ appearing in the summation in formula~\eqn{shift} suffices to 
gauge away the operators $H(\wt \nabla,Y)$. One might worry that this reintroduces new 
contributions to $Q_{22}$ at order zero and one in $\wt \nabla$, but we have as yet not used the freedom to choose
the first two terms in~\eqn{shift}. Clearly, when $U_M\neq 0$, we can choose $\varepsilon^{MN} U_M$ to ensure that $Q_{22}$
has no term linear in $\wt \nabla$. Finally, when $U_M$ is not a conformal Killing vector (notice that~\eqn{XXE} implies that $\varepsilon^{MN}$ is trace-free) we can try to use the first term in~\eqn{shift} to remove~$\Sigma$.
A generic choice of metric $G_{MN}$ will not admit conformal Killing vectors so we may safely\footnote{It is possible that $\Sigma$ can still be gauged away even if the metric $G_{MN}$ admits conformal Killing vectors $U^M=\nabla^M\Sigma$. We have not analyzed this
issue in detail, but it interesting to note that the condition $\nabla_{(M} U_{N)}\propto G_{MN}$ along with the weight condition~\eqn{weight} for $\Sigma$ implies that $\Sigma$ is an eigenstate of the quadratic Casimir of the triplet of operators~\eqn{sp2}. 
} 
pick a gauge for which $\Sigma =0$.
 
Thus, we arrive at our final solution for the quantum equations~\eqn{QQ}
\be\label{Qquf}
Q(G_{MN},A_M)=
\begin{pmatrix}
X^2 & X^M \wt \nabla_M +\frac{d+2}{2}\\[3mm] X^M \wt \nabla_M  +\frac{d+2}{2}\ &  
\wt \nabla^2 \end{pmatrix}\, .
\ee
It is parameterized, modulo diffeomorphisms and $SO(1,1)$ gauge transformations by a metric $G_{MN}$ and abelian gauge field $A_M$ 
subject to the closed homothety and transverse curvature requirements in equations~\eqn{key} and~\eqn{XF}, respectively. This is the gravity mulitplet of our model. It describes spacetime geometry but does not describe gravitational dynamics. From the tractor viewpoint, 
that requires coupling to scale. Or in other words, a dilaton. Therefore, we now describe the coupling of the gravity multiplet
to the dilaton multiplet.
\section{Main Results: Gravity}
\label{main}
In section~\ref{Scale Tractor} we saw that instead of formulating gravity in terms of an  Einstein--Hilbert action functional
depending on four-metrics, one could build  from the square of the scale tractor~$I^M$ 
an equivalent action depending on the scale (or dilaton)~$\sigma$ and a conformal class of four dimensional metrics~$[g_{\mu\nu}]$.
The operator $Q$ of the previous section depended on  (i) a six dimensional metric $G_{MN}$ with closed homothety
and (ii) a six dimensional vector $A_M$. Since the metric $G_{MN}$ encodes a four dimensional conformal class of metrics~$[g_{\mu\nu}]$
one can hope that the vector $A_M$ is somehow related to the scale tractor and so that a theory built from the operator $Q$
could amount to a tractor description of Einstein--Hilbert gravity. For this proposal to work, we still need to couple to a dilaton field, or in
other words scalar matter. From a two times physics perspective this coupling should respect the gauge symmetry~\eqn{Qs} as well as 
the~$\frak{sp}(2)$ gauge symmetry generated by the operators~$Q$. A coupling to scalars with exactly these symmetries has been computed by Bars using first quantized BRST techniques \cite{Bars:2006ir} and reads
\be\label{BRST}
S(Q,\Omega, \Theta,\Lambda,\Psi)=\scalebox{1.1}{$ \frac{2(d-1)}{d-2}$}\int d^{d+2}Y \sqrt{G} \, \big[\Omega\,  Q_{22} + \Theta\,  Q_{12} + \Lambda\,  Q_{11}\big]\Psi\, .
\ee
Our claim is that this action principle, along with the conditions~\eqn{QQ} on the operator $Q$ amounts to the tractor description 
of four dimensional Einstein--Hilbert gravity.

The action~\eqn{BRST} depends (from a six dimensional viewpoint) on an infinite set of fields through the operator $Q$. However
it also enjoys infinitely many local symmetries generated by an operator parameter $\epsilon$ as well as a local $\frak{sp}(2)$ invariance
with local parameters $(\lambda(Y),\theta(Y),\omega(Y))$
\bea
Q&\mapsto& Q + [\epsilon, Q]\, ,\nn\\[2mm]
\Psi&\mapsto& \Psi + \epsilon \Psi\, ,\nn\\[2mm]
\Omega&\mapsto& \Omega -\epsilon^\dagger \Omega - Q_{11}^\dagger \theta + [Q_{12}^\dagger+2] \omega \,  ,\nn\\[2mm]
\Theta&\mapsto & \Theta -\epsilon^\dagger \Theta +Q_{11}^\dagger \lambda - Q_{22}^\dagger \omega - 4 \, \theta\, ,\nn\\[2mm]
\Lambda &\mapsto & 
\Lambda -\epsilon^\dagger \Lambda + Q_{22}^\dagger \theta - [Q_{12}^\dagger -2]\lambda\, .\nn
\eea  
Here the dagger operation is the standard adjoint with respect to the six dimensional measure appearing in~\eqn{BRST}.
We are now ready to verify our claim that~\eqn{BRST} is the theory of gravity.

The first step is use the gauge freedom~$\epsilon$ to reach the gauge~\eqn{Qquf} for the operator $Q$. This yields a standard, generally covariant,  six dimensional action depending only on finitely many fields $(G_{MN},A_M,\Omega,\Theta,\Lambda)$
\be\label{action}
S={\scalebox{1.1}{$ \frac{2(d-1)}{d-2}$}} \int  d^{d+2}Y
\sqrt{G}~\Big[ \Omega\, \wt \nabla^2+\Theta \big(X^M\wt\nabla_M+\frac{d+2}{2}\big)+\Lambda X^2\Big]\Psi\, ,
\ee
with gauge invariance
\bea
 A_M&\mapsto&A_M+\nabla_M\alpha\, ,\nonumber\\[1mm]
 \Psi\ &\mapsto& \Psi -\alpha \, \Psi\, ,\nn\\
\Omega\ &\mapsto&\Omega+\alpha\, \Omega-X^2\theta-\big(X^M\wt\nabla_M+\frac{d+2}{2}-2\big)\omega\, ,\nonumber\\
\Theta\ &\mapsto&\Theta+\alpha\, \Theta+X^2\lambda-\wt\nabla^2\omega-4\, \theta\, ,\nonumber\\
\Lambda\ &\mapsto&\Lambda + \alpha\, \Lambda+\wt\nabla^2\theta +\big(X^M\wt\nabla_M+\frac{d+2}{2}+2\Big)\lambda\, .\label{6g}
\eea 

The action~\eqn{action} is four dimensional gravity wearing a six dimensional disguise. To disrobe it further, we use the $SO(1,1)$ gauge symmetry $\alpha$ to choose a gauge
 \be\label{gauge}
 X^M A_M=-w~~~~~~~\textrm{which implies}~~~~~X^N \nabla_N A_M=-A_M\,  .
 \ee
Here $w$ is an arbitrary real number. We could equally well have chosen $w=0$, but we prefer the above since it will imply the most general 
 assignments of tractor weights to the scalar fields. In any case,~$w$ will drop out at the end of our computation, and thereby serves as a check on our algebra. Notice that using~\eqn{XF}, the potential~$A_M$ now has weight $-1$ with respect to the weight operator~$X^M \nabla_M$. Note that the vector $A_M$ still enjoys residual abelian gauge transformations with weight zero gauge parameter $X^M \nabla_M \alpha=0$.
 
We now integrate out the Lagrange multipliers~$(\Theta,\Lambda)$ which imposes constraints
\be\nn
X^M \nabla_M \Psi = \big(w-\frac d2 -1\big)\Psi\, ,\qquad X^2 \, \Psi=0\, . 
\ee
Solving the latter constraint via
\be\nn
\Psi=\delta(X^2)\phi\, ,\qquad \phi\sim\phi + X^2 \chi\, ,
\ee
and comparing with~\eqn{cone} and~\eqn{w}, we see that $\phi$ is a weight~$w-\frac d2 +1$ tractor scalar.

There is still the freedom using the gauge parameter~$\omega$ to gauge away $\Omega$ save for gauge transformations $\omega$ in the kernel of~$X^M \nabla_M +w +\frac d2 -1$. Hence all that remains is the part of $\Omega$ of weight $-w-\frac d2 +1$. The remaining field content along with their  weights are summarized in 
 the following table\\~\\
 \begin{center}
 \begin{tabular}{c|c}
~~~~~Field~~~~ ~&~~~~~ Weight~~~~~\\\hline\\[-3mm]
$\Omega$&$-w-\frac d 2 +1$\\[2mm] $\phi$&$\phantom{-}w-\frac d 2 +1$ \\[2mm]$A_M$&$-1$
\end{tabular}
\end{center}
Integrating by parts to ensure no derivatives act on the delta function in $\Psi$, the action now takes the extremely simple form\be\label{action2}
S={\scalebox{1.1}{$ \frac{2(d-1)}{d-2}$}} \int d^{d+2}Y \sqrt{G}\, \delta(X^2)~ T\, ,
\ee 
where 
\be\label{T0}
T=\phi (\nabla^M-A^M)(\nabla_M-A_M) \Omega\, .
\ee
Since $T\sim T+X^2 \, U$, it is a tractor scalar with weight $-d$ (see the above table). We would like to express 
the action~\eqn{action2} as a four dimensional integral over tractor-valued objects\footnote{Bars handles delta-function valued ambient
space integrals by developing a calculus for derivative of delta functions~\cite{Bars:2008sz}. The simple tractor analysis given here, obviates the need for such methods. }.
To that end we need to express~\eqn{T0}
in terms of ambient tractor operators: Using the ambient expression~\eqn{DDD} for the Thomas $D$-operator,  
we easily derive the following ambient tractor identities
\bea
\Delta\Omega -2A^M\nabla_M\Omega
=\frac 1w
A^MD_M\Omega\, ,\nn\\[2mm]
\nabla^M A_M = \frac{1}{d-2}\, D_M A^M\, .\quad
\eea
(There is no pole at $w=0$ in the first identity, as can be easily verified by using the four dimensional component expression~\eqn{thomasd}
for the Thomas $D$-operator.) Hence
\be\nn
T=\phi\Big( \frac 1 w A^M D_M-\frac{1}{d-2}(D_M A^M)+A^2 \Big)\Omega\, .
\ee
The beauty of this expression is that $\delta(X^2) T$ now only depends on equivalence classes $A_M\sim A_M + X^2 B_M$, 
$\Omega \sim \Omega + X^2\,  \Xi$. Therefore {\it all} fields are now tractor valued. Hence we may replace the ambient space integral~\eqn{action2}, with a four dimensional integral depending on tractors $(\phi,\Omega,A_M)$
\be\label{action3}
S={\scalebox{1.1}{$ \frac{2(d-1)}{d-2}$}}\int d^dx\sqrt{-g}\, \phi\Big[ \frac 1 w A^M D_M-\frac{1}{d-2}(D_M A^M)+A^2 \Big]\Omega\, .
\ee
Note that the integrand has weight $-d$ while the metric determinant has weight $d$ under Weyl transformations so this action 
principle is now manifestly Weyl invariant. Our claim is now that this tractor action is equivalent to the formulation of the Einstein--Hilbert 
action in terms of the square of the scale tractor~\eqn{I^2 action for gravity}.

To verify our final claim we must examine the remaining $SO(1,1)$ gauge symmetry
\bea
A_M&\mapsto&A_M+\frac{1}{d-2}D_M \alpha \, ,\nn\\[2mm]
\Omega&\mapsto& \ \Omega \ +\ \alpha \, \Omega\,  ,\nn\\[2mm]
\phi&\mapsto &~ \phi ~ - ~  \alpha\, \phi\, ,
\eea
where the gauge parameter $\alpha$ is a weight zero tractor scalar. 
Notice that the gauge transformation of $A_M$ respects the condition $X^M A_M = -w$.
Now observe that the action depends only algebraically on the $SO(1,1)$ 
gauge field~$A_M$ and  the pair of fields $(\phi,\Omega)$ form a doublet under this symmetry.
Hence, we expect that upon integrating out $A_M$, only the gauge invariant combination $\phi\, \Omega$
should survive. This computation can be performed either using component expressions for the tractor 
quantities in~\eqn{action3} or directly using tractors. In components, one finds that the bottom slot $A^-$ of 
the gauge field decouples completely from the action and that integrating out the middle slot of $A_M$ sets it equal to 
the $SO(1,1)$ current $\frac12 \nabla_m\log(\Omega/\phi)$. This yields the four dimensional action for a conformally improved scalar field
\be\nn
S={\scalebox{1.1}{$ \frac{2(d-1)}{d-2}$}} \int d^d x\sqrt{-g}~\varphi\Big[ \Delta-\frac{d-2}{2}~\Rho \Big]\varphi\, ,
\ee
where $\varphi$ is the weight $1-\frac d 2$ scalar field defined by 
$$
\varphi^2=\phi\, \Omega\, .
$$
In other words it is the \emph{dilaton}.
Using the relationship between the dilaton and scale, $\varphi=\sigma^{1-\frac d2}$, we obtain as explained in section~\ref{Scale Tractor} the tractor version
of the Einstein--Hilbert action in terms of the square of the scale tractor
\be\label{II}
S=\scalebox{1.1}{$\frac{d(d-1)}{2}$}\int d^dx\frac{\sqrt{-g}}{\sigma^d} \, I^MI_M\, .
\ee
This completes our demonstration that the $\frak{sp}(2)$ invariant theory~\eqn{BRST} amounts to a theory of four dimensional gravity.
We now turn to implications of our results.

\section{Conclusions and Outlook}\label{conc}
In this article we formulated the Einstein--Hilbert action as a trace 
\be\label{quantum}
S={\rm tr} \ Q\, P
\ee
over quantum mechanical operators $Q$ (as in~\eqn{Qquf}) and
$$
P=\begin{pmatrix}\ \, |\Psi\rangle\langle\Lambda|& \frac12|\Psi\rangle\langle\Theta|\\[2mm] \frac12|\Psi\rangle\langle\Theta|&\ \, |\Psi\rangle\langle\Omega|\end{pmatrix}\, .
$$
In this formulation, second quantization amounts to integrating over the space of operators $Q$ and $P$ in the path integral.
This leads one to wonder whether quantum field theory effects, such as Weyl anomalies, can be understood from this six-dimensional quantum mechanical picture. An advantage of this two times approach is that it formulates gravity in terms of a very limited field content: the three components of $Q$ viewed as functions of a twelve dimensional phase space. Weyl and diffeomorphism symmetries are neatly encoded in the algebra~\eqn{QQ} and its gauge invariance~\eqn{Qs}. A pressing question therefore is to compute anomalies in the
$\frak{sp}(2)$ symmetry.

Another benefit of the two times starting point~\eqn{BRST} is that it yields a new tractor formulation of the conformally Einstein
condition (see the action~\eqn{action3}). At the very least, this should have implications for conformal geometry; the triplet 
of tractor fields $(\phi,\Omega,A_M)$ underly the scale tractor~$I^M$. This observation deserves further investigation.

Another interesting avenue for further research is whether there exists a framelike formulation of two times physics. This is
based on the simple observation that the operator~\eqn{Qquf} can be factorized as
\be\nn
Q=\left[
\begin{pmatrix}
X_M \\[3mm] \wt \nabla_M
\end{pmatrix}
\begin{pmatrix}
X^M & \wt \nabla^M
\end{pmatrix}
\right]_W\, .
\ee
The operator $V^M_i=(X^M\ \ \wt \nabla^M)$ can then be interpreted as a two times frame field,
so one could try to impose the Howe dual pair~\eqn{how} decomposition as  equations
of motion for fundamental fields $V^M_i$. This might be particularly interesting when one considers
the interpretation of the infinite tower of six dimensional auxiliary fields appearing in the parent action~\eqn{BRST}.
In particular, one wonders whether these fields solve the problem posed, and partially solved in~\cite{MAV}, of finding 
an unfolding of the full nonlinear Einstein's equations. The relation between these two approaches may be clearer
in a framelike formulation, since (unlike unfolding constructions)  two times models are typically constructed in a metric formulation.

Finally, a gravitational two times action principle that simultaneously incorporates the benefits of both actions~\eqn{BRST} and~\eqn{barsactiongrav}---namely producing the $\frak{sp}(2)$ algebra as equations of motion while maintaining manifest $\frak{sp}(2)$ symmetry---would be very desirable. In fact, once we understand that our work implies that the coupling of the gravity multiplet (built from $\frak{sp}(2)$ generators) to scalars really amounts to a gravity-dilaton coupling, then we can identify yet another action principle proposed by Bars
as a candidate model for cosmological four dimensional Einstein gravity. Bars' proposal is to produce the equations of motion
for the operator $Q$ from a Chern--Simons action~\cite{Bars:2001ma}
$$
S_{\rm CS}=\int \big[Q\star Q + Q\star Q\star Q\big]\, ,
$$ 
(where the Moyal star product~$\star$ is employed to produce operator equations of motion from phase space valued fields).
Hence the sum of this action plus the BRST action $S_{\rm BRST}$ in~\eqn{BRST}
\be\label{tot}
S=S_{\rm CS} + \lambda S_{\rm BRST}\, ,
\ee
deforms the $\frak{sp}(2)$ relations by dilaton dependent terms (see~\cite{Bars:2001ma} for explicit formul\ae). A simple conjecture, therefore, is 
that these produce the cosmological constant coupling missing from the action~\eqn{BRST}. In particular, the relative coefficient $\lambda$
in the total action~\eqn{tot} could be identified with the cosmological constant.

\section*{Acknowledgements}
It is a pleasure to thank Itzhak Bars, Nicolas Boulanger, David Cherney, Tom Curtright, Olindo Corradini, Claudia de Rham, Rod Gover, Maxim Grigoriev, Djordje Minic, Bruno Nachtergaele, Andy Port, Abrar Shaukat, Per Sundell, Misha Vasiliev and Steven Weinberg for discussions. R.B. thanks the U.C. Davis Department of Mathematics for warm hospitality and the INFN, Sezione di Bologna, and the Universit\`a di Bologna Marco Polo program for financial support.

\appendix
\section{An Alternative Six Dimensional Formulation of Gravity}\label{alt}
In \cite{Bars:2008sz} Bars proposed the following six dimensional field theory model for gravity coupled to scalar field
\bea\label{barsactiongrav}
S&=&-\frac 1 2\int d^{d+2}Y\sqrt{G}\, \Big[\delta(W)\Big(R(G)\varphi^2
+\alpha\, (\nabla\varphi)^2-\lambda\varphi^{\frac{2d}{d-2}}\Big)~~~~~~~~~~~\nonumber\\[1mm]
&&~~~~~~~~~~~~~~~~~~~~-\delta'(W)\Big((\Delta W-4)\ \varphi^2-\nabla_M W\ \nabla^M\varphi^2\Big)\Big]\, ,
\eea
with $\alpha=\frac{4(d-1)}{d-2}$ and  for some $\lambda$ playing the {\it r\^ole} of the cosmological constant.  A distinguishing feature of this action is that the homothetic condition and the weight condition on $\varphi$ follow from its equations of motion; they indeed arise from the field equations for $G_{MN}$ and $\varphi$ instead of requiring closure of the $\frak{sp}$(2) algebra.  Partially solving those equations, one obtains the following set of relations
$$W=X^2,~~~  G_{MN}=\nabla_MX_N,~ ~~ X^M\nabla_M \varphi=\big(1-\frac d 2 \big)\varphi~.$$ Plugging these back in~\eqref{barsactiongrav} we get the following model
\bea\label{bars2tgravity}
S&=&-\frac 1 2\int d^{d+2}Y\sqrt{G}\, \delta(X^2)\Big[R(G)\varphi^2
-\alpha\, \varphi\Delta\varphi-\lambda\, \varphi^{\frac{2d}{d-2}}\Big]\, .
\eea
Now note that, introducing the scale tractor $I^M$ constructed from $\sigma=\varphi^{\frac{2}{2-d}}$ in the usual way (see section \ref{Scale Tractor}), the action \eqref{bars2tgravity} becomes
\bea
S&=&-\frac 1 2\int d^{d+2}Y\sqrt{G}\, \delta(X^2) \Big[ R(G)\varphi^2
+\frac{\alpha}{\sigma}\, \varphi\,  I^M D_M \, \varphi-\lambda\, \varphi^{\frac{2d}{d-2}} \Big]\, ,\nn
\eea
that in turn, by using the relation $I^M D_M \sigma^k= k(d+k-1)\sigma^{k-1}I^2$, can be rewritten as
 \bea\label{inalction}
S&=&-\frac 1 2\int d^{d+2}Y\sqrt{G}\, \delta(X^2)~\frac{1}{\sigma^d}\Big[ R(G)\sigma^{2}
-d(d-1)I^2-\lambda~\Big]~.
\eea
Let us observe at this point that, as was shown by Fefferman and Graham in~\cite{FG}, a conformal class of $d$-dimensional metrics $[g_{\mu\nu}]$ determines a Ricci flat ambient space  if~$d$ is odd, and a Ricci flat ambient space modulo~$(X^2)^{\frac{d-2}{4}}$. Hence, since
the action~\eqn{inalction} depends only on the conformal class of metrics~$[g_{\mu\nu}]$ and includes the delta function~$\delta(X^2)$, 
we can set to zero the curvature term in~\eqref{inalction}. In fact, another way to see this, is that we could have chosen a gauge
in section~\ref{2TP} where $\Sigma=R(G)$.

Now that the model is completely written in terms of tractor objects it may be directly written in four dimensional language as
\bea\label{finalction}
S=\scalebox{1.1}{$\frac{d(d-1)}{2}$}\int d^dx\frac{\sqrt{-g}}{\sigma^d}\, \Big[ 
I^MI_M+\frac{\lambda}{d(d-1)}\Big]~.
\eea
When~$\lambda=0$, this model coincides with~\eqn{II} demonstrating the equivalence of these two models in that case.
The formulation~\eqn{barsactiongrav} has the advantage that it includes a cosmological constant and partially imposes the relations~\eqn{QQ} as equations of motion coming from a variational principle. Its disadvantage is that the manifest $\frak{sp}(2)$ symmetry
is lost. In our conclusions we speculated that a third model proposed by Bars incorporates the best features of both models~\eqn{BRST}
and~\eqn{barsactiongrav}.

\newpage
\end{document}